\documentclass[11pt,a4paper]{article}
\usepackage{bm, amssymb,pifont,cancel,amsmath,comment,authblk}
\usepackage[dvips]{graphicx}
\usepackage[colorlinks=true,link color=blue,cite color=blue,urlcolor=black]{hyperref}
\usepackage{braket}

\newcommand{\beq}{\begin{eqnarray}}
\newcommand{\eeq}{\end{eqnarray}}
\newcommand{\p}{\partial}

\newcommand{\hs}[1]{\hspace{#1 mm}}
\newcommand{\bpm}{\begin{pmatrix}}
\newcommand{\epm}{\end{pmatrix}}

\newcommand{\C}{\mathbb{C}}

\newcommand{\ba}{\left(\begin{array}}
\newcommand{\ea}{\end{array} \right)}

\begin{document}

\begin{center}
{\Large{\bf{Ghostbusters in higher derivative supersymmetric theories:\\
who is afraid of propagating auxiliary fields?}}}
\vskip 6pt
\large{Toshiaki Fujimori$^{1,2}$}{\renewcommand{\thefootnote}{\fnsymbol{footnote}}},
\large{Muneto Nitta$^{1,2}$}{\renewcommand{\thefootnote}{\fnsymbol{footnote}}} and \large{Yusuke Yamada$^{2,3}$}{\renewcommand{\thefootnote}{\fnsymbol{footnote}}}\\
\vskip 4pt
{\small{\it 
$^1$ Department of Physics, Keio University,}}\\
{\small{\it  Hiyoshi 4-1-1, Yokohama, Kanagawa 223-8521, Japan}}\\
\vskip 1.0em
{\small{\it 
$^2$ Research and Education Center for Natural Sciences,
Keio University,}}\\
{\small{\it Hiyoshi 4-1-1, Yokohama, Kanagawa 223-8521, Japan}}
\\
\vskip 1.0em
{\small{\it 
$^3$ Department of Physics, Keio University,}}\\
{\small{\it Hiyoshi 3-14-1, Yokohama, Kanagawa 223-8522, Japan}}
\end{center}
\begin{abstract}

We present for the first time  
a ghost-free higher-derivative chiral model 
with a propagating auxiliary F-term field (highest component of 
the chiral multiplet). 
We obtain this model by removing a ghost 
in a higher derivative chiral model,  
with Higgsing it in terms of an auxiliary vector superfield.  
Depending on the sign of the quadratic derivative term of 
the chiral superfield,
the model contains two ghost free branches of the parameter regions.
We find that supersymmetry is spontaneously broken in 
one branch while it is preserved in the other branch. 
As a consequence of dynamical F-term field, 
a conserved U(1) charge corresponding to 
the number density of $F$ appears, 
which can be regarded as a generalization of the R-symmetry. 
\end{abstract}
\section{Introduction}
Supersymmetry (SUSY) is an interesting extension beyond the standard model of particle physics (SM) from various viewpoints. The SUSY extension of SM has dark matter candidates and a nice property controlling quantum corrections to the Higgs mass. It is also known that local SUSY, called supergravity (SUGRA), is the effective theory of superstring, which is a possible quantum gravity theory. Once we start with SUGRA, the renormalizability of a theory disappears and we cannot control such terms. Although such a non-renormalizability may be restored in the UV complete theory, we need to consider higher-order terms in the effective theory. If this is the case, higher-order derivative interactions would appear as the non-derivative higher-order terms exist.

It has been well known that higher-order derivative interactions may lead to the so-called Ostrogradski instability 
(see Ref.~\cite{Woodard:2006nt} as a review) 
because
the energy of such a system cannot be bounded. Therefore, it is important to specify
a class of ghost-free higher-derivative interactions. For non-SUSY scalar-tensor theory, Horndeski found a class of higher-derivative Lagrangian without ghosts~\cite{Horndeski:1974wa} (see also Refs.~\cite{Kobayashi:2011nu,Charmousis:2011bf}).

In SUSY cases, 
ghost-free higher-derivative interactions have been studied 
in various contexts.
The properties of such higher-derivative terms have been studied from theoretical and phenomenological viewpoints.
For vector and real linear superfield, Dirac-Born-Infeld action was constructed in Refs~\cite{Bagger:1996wp,Bagger:1997pi,Rocek:1997hi}. Their generalizations to SUGRA have also been discussed in Refs.~\cite{Kuzenko:2002vk,Kuzenko:2005wh,Farakos:2012je,Koehn:2012ar,Abe:2015nxa,Aoki:2016cnw}. 
For a single chiral superfield, some types of ghost-free higher-derivative terms have been known~\cite{Khoury:2010gb,Farakos:2013zya}. 
Their cosmological applications, especially to the inflationary universe, have also been discussed in Refs.~\cite{Sasaki:2012ka,Koehn:2012np,Gwyn:2014wna,Aoki:2014pna,Aoki:2015eba,Abe:2015fha}. 
BPS states in supersymmetric
higher derivative chiral models were also studied,
such as
baby-Skyrmions 
\cite{Adam:2011hj,Adam:2013awa,Nitta:2014pwa,Bolognesi:2014ova,Queiruga:2016jqu} 
and 
various BPS topological solitons in the higher derivative 
chiral model \cite{Nitta:2014pwa}
coupled with gauge theory \cite{Nitta:2015uba}.
The other examples are 
low-energy effective action of supersymmetric QCD 
\cite{Buchbinder:1994iw,Banin:2006db},  
the Wess-Zumino-Witten (WZW) term 
in the supersymmetric chiral Lagrangian  
\cite{Nemeschansky:1984cd,Gates:1995fx,Gates:2000rp,Nitta:2001rh},
supersymmetric nonlinear realizations of spontaneously 
broken global symmetry \cite{Nitta:2014fca},
the low-energy effective theory on BPS solitons \cite{Eto:2012qda},
k-field theories \cite{Adam:2011gc,Adam:2012md},
the Faddeev-Skyrme model 
\cite{Freyhult:2003zb,Bergshoeff:1984wb},
and 
(pure) Skyrme model \cite{Gudnason:2015ryh}. The low-energy effecitive action with loop effects has infinite numbers of derivative operators~\cite{Gomes:2009ev,Gama:2011ws,Gama:2014fca}.
A ghost free higher derivative theory is also studied 
recently in the framework of non-local field theories \cite{Kimura:2016irk}.

In general,
higher derivative terms in supersymmetric theories 
suffer from another problem specific for 
supersymmetry, namely 
the auxiliary field problem:
the auxiliary field $F$ (F-term) of a chiral superfield may propagate 
because of space-time derivative terms acting on it, 
and consequently it cannot be eliminated algebraically. 
This problem was seriously recognized 
\cite{Gates:1995fx,Gates:2000rp} 
for the WZW term. 
As systematically studied in Ref.~\cite{Khoury:2010gb}, 
the models mentioned in the above paragraph 
are free from this problem 
(except for that in Ref.~\cite{Bergshoeff:1984wb}).
In Ref.~\cite{Antoniadis:2007xc} (see also \cite{Dudas:2015vka}), 
a very important model with 
a single chiral superfield with higher derivative term was
studied, in which the auxiliary field $F$ of the chiral superfield 
in fact becomes dynamical.
In this model, a ghost chiral superfield is induced due to 
the higher derivative term.
Because of this model, 
it is widely believed 
without any proof 
that the above mentioned two problems are related: 
when an auxiliary field becomes dynamical, 
a ghost should be present and the theory is pathological.

In this paper, we present a first counterexample to such a conjecture, {\it i.e.}
a ghost-free higher-derivative chiral model 
with a propagating auxiliary field $F$.\footnote{
The propagating auxiliary fields have been discussed in the context of the higher-curvature SUGRA model~\cite{Cecotti:1987sa,Farakos:2013cqa,Ketov:2013sfa,Ferrara:2013wka,Ketov:2013dfa}, in which
 the kinetic term of 
auxiliary fields in the gravity multiplet exists 
due to higher-derivative terms.  
}
We achieve this by removing a ghost in the higher derivative 
chiral model in Ref.~\cite{Antoniadis:2007xc}; 
the ghost is Higgsed away by 
a non-dynamical auxiliary vector superfield $V$ associated with a 
U(1) gauge symmetry.\footnote{
Such auxiliary vector superfields were used to formulate  
supersymmetric nonlinear sigma models 
in terms of gauge theories \cite{D'Adda:1978uc,Higashijima:1999ki}. 
When $V$ has no kinetic term, its vector component plays a role of an auxiliary field, and gaugino and auxiliary components behave as Lagrange multiplier fields, which give rise to constraints on the coupled superfields. Also, we need to fix the gauge degree of freedom.  
}
As we will show, with appropriate couplings of chiral and gauge superfields, the ghost degrees of freedom can be removed thanks to the constraints and gauge degree of freedom. 
Depending on the sign of the quadratic derivative term of 
the chiral superfield,
the model contains two ghost free branches of the parameter regions. 
We find that supersymmetry is spontaneously broken in 
one branch while it is preserved in the other branches,
which are totally unexpected in the original Lagrangian.
The auxiliary field $F$ in the original chiral superfield 
is now in the lowest component of 
another chiral superfield after ``unfolding'' 
the higher derivative term to two chiral superfields 
with second derivative terms. 
As a consequence, a dynamical auxiliary field 
allows an unexpected U(1) symmetry 
and associated conserved charge,
which are not manifest in the original Lagrangian. 
This can be regarded as a generalization of the R-symmetry, 
so we may call it an R'-symmetry.

The remaining part is organized as follows. In Sec.~\ref{rev}, we review a SUSY model with a higher-derivative term, which produces a ghost mode. We also find that an auxiliary field of a chiral superfield obtain its kinetic term, which is an additional mode due to a SUSY higher derivative. We extend the higher-derivative action to that coupled to a gauge superfield in Sec.~\ref{exo}. We show how the ghost can be removed by such an extension, and find that, even after eliminating a ghost superfield, the superfield originated from a dynamical auxiliary field remains in the resultant system. In Sec.~\ref{DF-term}, we briefly discuss some features of the resultant system. Finally, we conclude in Sec.~\ref{concl}.

\section{SUSY higher-derivative ghost}\label{rev}
In this section, we review a model with a higher-derivative term inducing ghost modes. 
As an illustration of our new proposal, 
we will extend the model to that with a gauge superfield in the next section. 
Here, let us consider the following higher-derivative Lagrangian,
\begin{align}
{\cal L}=\int d^4\theta\left[\Phi\overline{\Phi}+\frac{\alpha}{\Lambda^2} D^2\Phi\overline{D}^2\overline{\Phi}\right]\label{GA}
\end{align}
where $\Lambda$ is a real constant of mass dimension one, 
and $\alpha$ is a dimensionless real parameter. 
A chiral superfield $\Phi$ is defined as
\begin{align}
\Phi & =\phi(y)+\sqrt{2}\theta\psi(y)+\theta\theta F(y)\nonumber\\
& = \phi(x)+\sqrt{2}\theta\psi(x)+\theta\theta F(x)+i\theta \sigma^m\overline{\theta}\partial_m\phi(x)-\frac{i}{\sqrt{2}}\theta\theta\partial_m\psi(x)\sigma^m\overline{\theta}+\frac{1}{4}\theta\theta\overline{\theta}\overline{\theta}\Box\phi(x),
\end{align}
where $y^m=x^m+i\theta\sigma^m\overline{\theta}$. 
In terms of the component fields, 
the explicit form of the Lagrangian is given by
\beq
\mathcal L = - \p_\mu \phi \p^\mu \bar{\phi} - i \bar{\psi} \bar \sigma^\mu \p_\mu \psi + |F|^2 
+ \frac{16 \alpha}{\Lambda^2} \left( - \p_\mu F \p^\mu \bar{F} - i \Box \bar{\psi} \bar{\sigma}^\mu \p_\mu \psi + |\Box \phi|^2 \right). 
\label{component}
\eeq
This Lagrangian has the U(1)$^3$ symmetry corresponding to the phases of $(\phi, \psi, F)$. 
The rotation of overall phase is the U(1) symmetry which commutes with SUSY, 
whereas the R-symmetry is the phase rotation with the charges $(0,1,2)$. 
The other U(1) symmetry, which we call the R'-symmetry, 
is the phase rotation of $F$, whose conserved charge is 
non-trivial due to the presence of the kinetic term of $F$. 
Note that this symmetry exists in theories without the dynamical F-term field
but its conserved charge vanishes on-shell. 

With the technique called ``unfolding"~\cite{Antoniadis:2007xc,Dudas:2015vka}, 
we can rewrite this Lagrangian into that without higher-derivative terms as follows: 
Using a Lagrange multiplier chiral superfield $\Phi_1$, 
the Lagrangian~(\ref{GA}) can be rewritten as
\begin{align}
{\cal L} = \int d^4 \theta 
\left[ \Phi \overline{\Phi} + \alpha \Phi_2 \overline{\Phi}_2 \right] + 
\left\{\frac{\Lambda}{4} \int d^2\theta \, \Phi_1 \left( \Phi_2 - \frac{1}{\Lambda} \overline{D}^2 \overline{\Phi} \right) + {\rm h.c.} \right\}, \label{GA2}
\end{align}
where $\Phi_2$ is a chiral superfield. 
For later convenience, we have chosen the normalization of $\Phi_1$ 
so that the overall coefficient of the superpotential becomes $\Lambda/4$. 
The variation with respect to $\Phi_1$ gives the constraint 
$\Phi_2=\frac{1}{\Lambda}\overline{D}^2\overline{\Phi}$, 
which reproduces the original Lagrangian~(\ref{GA}). 
Instead, if we use the following identity, 
\begin{align}
\int d^2\theta \, \overline{D}^2(\Phi_1\overline{\Phi})+{\rm h.c.}=\int d^4\theta \left(-4\Phi_1\overline{\Phi}+{\rm h.c.}\right) ~~~(\mbox{up to total derivative terms}),
\end{align}
the Lagrangian~(\ref{GA2}) becomes
\begin{align}
{\cal L} =& 
\int d^4 \theta \Big[ \Phi \overline{\Phi} + \left( \Phi_1 \overline{\Phi} + {\rm h.c.} \right) + \alpha\Phi_2 \overline{\Phi}_2 \Big] 
+ \left\{ \frac{\Lambda}{4} \int d^2 \theta \left( \Phi_1 \Phi_2 \right) + {\rm h.c.} \right\} \nonumber\\
=& \int d^4 \theta \Big[ \left|\Phi + \Phi_1 \right|^2 - \Phi_1 \overline{\Phi}_1 + \alpha \Phi_2 \overline{\Phi}_2 \Big]
+ \left\{ \frac{\Lambda}{4} \int d^2 \theta (\Phi_1 \Phi_2 ) + {\rm h.c.} \right\} \nonumber \\
=& \int d^4 \theta \Big[ \tilde{\Phi}\overline{\tilde{\Phi}} - \Phi_1 \overline{\Phi}_1 + \alpha \Phi_2 \overline{\Phi}_2 \Big]
+ \left\{ \frac{\Lambda}{4} \int d^2 \theta (\Phi_1 \Phi_2 ) + {\rm h.c.} \right\}, \label{GA3}
\end{align}
where 
\begin{align}
\tilde{\Phi} \equiv \Phi + \Phi_1.
\end{align} 
From Eq.\,(\ref{GA3}), 
we find that $\Phi_1$ has a negative definite kinetic coefficient, 
that is, $\Phi_1$ is a ghost superfield. 
Depending on the sign of $\alpha$, $\Phi_2$ is either a ghost or regular superfield.

Let us focus on $\Phi_2$. 
In our discussion above, 
$\Phi_2$ came from $\overline{D}^2\overline{\Phi}$, 
whose lowest component is $F^\Phi$. 
Indeed, the component expression of the second term in the action~(\ref{GA}) is given by
\begin{align}
\int d^4 \theta \left( D^2\Phi \overline{D}^2\overline{\Phi} \right) =&
16 \Big( F \Box \overline{F} + \Box \phi \Box \overline{\phi} - i \Box \overline{\psi} \, \overline{\sigma}^n \partial_n \psi \Big).
\end{align}
Thus, we can identify $\Phi_2$ as the ``dynamical" F-component 
due to the SUSY higher-derivative contribution. 
It is important to note again that 
the presence of the higher-derivative term here is problematic 
since at least one ghost mode appears, 
irrespective of the value of $\alpha$.

\section{Removing ghost and dynamical F-term}\label{exo}

\subsection{Gauged model}
In this section, we discuss a possible modification of the higher-derivative system. 
A gauge symmetry is an important notion to remove some degree of freedom. 
We consider the case that the chiral superfield $\Phi$ is gauged under a U(1) symmetry. 
We introduce a gauge superfield $V$ for the U(1) gauge symmetry under which 
the superfields $\Phi$ and $V$ transform as
\begin{align}
\Phi \to& \Phi e^{-2i\Lambda},\\
V \to& V + i ( \Lambda - \overline{\Lambda}),
\end{align} 
where $\Lambda$ is a gauge parameter chiral superfield. 
The component expression of $V$ in Wess-Zumino gauge is
\begin{align}
V=-\theta\sigma^m\overline{\theta}v_m(x)+i\theta\theta\overline{\theta}\overline{\lambda}(x)-i\overline{\theta}\overline{\theta}\theta\lambda(x)+\frac{1}{2}\theta\theta\overline{\theta}\overline{\theta}{\cal D}(x).\label{VWZ}
\end{align}
The U(1) invariant extension of the Lagrangian~(\ref{GA}) is given by
\begin{align}
{\cal L}'=\int d^4 \theta \bigg[ \Phi e^{2V} \overline{\Phi} 
+ \frac{\alpha}{\Lambda^2} \Big( D^2( \Phi e^{2V}) \Big) e^{-2V} \Big( \overline{D}^2 ( \overline{\Phi} e^{2V} ) \Big) + 2 C V \bigg], \label{NA}
\end{align}
where we have introduced a possible Fayet-Iliopoulos (FI) parameter $C$. 
The higher-derivative superfield $\overline{D}^2 (\overline{\Phi} e^{2V})$ is a chiral superfield, 
whose component is given by
\begin{align}
\overline{D}^2 (\overline{\Phi} e^{2V}) =4 \Big[ -\overline{F}(y) + \sqrt{2} \theta \left( i \sigma^{\mu} D_{\mu} \overline{\psi}(y) - \sqrt{2} \lambda \overline{\phi}\right) - \theta \theta \left(\Box \overline{\phi}(y) + {\cal D}(y) \overline{\phi}(y) - \sqrt{2} i \overline{\lambda}(y) \overline{\psi}(y) \right) \Big],
\end{align}
where the covariant derivatives are 
$D_{\mu} \overline{\psi}=\partial_{\mu} \overline{\psi}-iv_{\mu} \overline{\psi}$, 
and $\Box\overline{\phi}=D^{\mu} D_{\mu} \overline{\phi}$. 
Rescaling $V \rightarrow g V$ and taking the limit of $g\to 0$, 
we obtain the component of $\overline{D}^2 \overline{\Phi}$.

\subsection{Component expression}
First, we illustrate how the ghost mode in the Lagrangian~(\ref{GA}) 
can be removed by the extension~(\ref{NA}).  
For simplicity, we focus on the bosonic part of the Lagrangian~(\ref{NA}), 
which is given by
\begin{align}
{\cal L}'|_B = - D_\mu \phi D^\mu \overline{\phi} + |F|^2 + {\cal D} \left( |\phi|^2 + C \right) 
+ \frac{16\alpha}{\Lambda^2} \Big[ - D_\mu F D^\mu \overline{F} + |\Box \phi + {\cal D} \phi|^2 - {\cal D} |F|^2 \Big], \label{BC}
\end{align}
where we have used the component expression of $V$ in the Wess-Zumino gauge~(\ref{VWZ}), 
$D_\mu F=\partial_\mu F + i A_\mu F$ and $\Box \phi=D^\mu D_\mu\phi$. 
To extract the ghost mode, we use the following trick: 
The Lagrangian~(\ref{BC}) can be rewritten as
\begin{align}
{\cal L}'|_B=& - D_\mu \phi D^\mu \overline{\phi} + |F|^2 + {\cal D} \left( |\phi|^2 + C \right) 
+ \frac{16\alpha}{\Lambda^2} \Big[ - D_\mu F D^\mu \overline{F} - {\cal D} |F|^2 \Big] \nonumber\\
& -\frac{\Lambda^2}{16\alpha} |\phi_1|^2 + \Big[ \overline{\phi_1} (\Box \phi + {\cal D}\phi) + {\rm h.c.} \Big], \label{BC2}
\end{align}
where $\phi_1$ is a scalar field with the same U(1) charge as $\phi$. 
It can be easily shown that the variation of $\phi_1$ reproduces the Lagrangian~(\ref{BC}). 
Here, we perform partial integration for the terms on the second line, which gives
\begin{align}
{\cal L}'|_B =& - D_\mu\phi D^\mu \overline{\phi} - \left\{D_\mu \overline{\phi}_1 D^\mu\phi+{\rm h.c.} \right\} - \frac{16\alpha}{\Lambda^2} D_\mu F D^\mu\overline{F} \nonumber\\
& + {\cal D}\left[|\phi|^2 + \left\{\phi \overline{\phi}_1 + {\rm h.c.} \right\} + \frac{16\alpha}{\Lambda^2} |F|^2 + C \right] 
+ |F|^2 - \frac{\Lambda^2}{16\alpha}|\phi_1|^2.
\end{align}
We find that the determinant of the kinetic coefficient matrix of $\phi$ and $\phi_1$ is negative
as with the case of the ungauged model.
However, we also find that ${\cal D}$ appears only linearly, 
and its E.O.M. gives a constraint on the scalar fields. 
In addition, we have the U(1) gauge symmetry which implies that 
there is a redundancy in the description of our model. 
Therefore, we can remove one complex scalar from the system 
by solving the constraint from ${\cal D}$, 
and eliminating the auxiliary vector field $A_\mu$. 
As we will see below, such a procedure can be simplified by using the superfield formalism. 
In the next subsection, we show by solving the E.O.M for the auxiliary vector superfield 
that the apparent ghost can be eliminated by the gauge symmetry. 

\subsection{Superfield calculation}

\subsubsection{The first model}
In this subsection, we show that the ghost mode can be gauged away
by eliminating the auxiliary vector superfield. 
As with the case of the original model~(\ref{GA2}), 
we introduce chiral superfields $\Phi_1$ and $\Phi_2$. 
In this case, they should have U(1) charges so that they transform
\begin{align}
\Phi_1 \to& \Phi_1 e^{-2i\Lambda},\\
\Phi_2 \to& \Phi_2 e^{2i\Lambda}.
\end{align} 
Using these superfields, we can construct the gauged version of the Lagrangian~(\ref{GA2}),
\begin{align}
{\cal L}=& \int d^4\theta \Big[ \Phi e^{2V} \overline{\Phi} + \alpha \Phi_2 e^{-2V} \overline{\Phi}_2 + 2 C V \Big] 
+ \left[ \frac{\Lambda}{4} \int d^2\theta \, \Phi_1 \left\{ \Phi_2-\frac{1}{\Lambda} \overline{D}^2 \Big( \overline{\Phi} e^{2V} \Big) \right\} + {\rm h.c.} \right] \nonumber \\
=&\int d^4\theta \left[ \tilde{\Phi} e^{2V}\overline{\tilde{\Phi}} - \Phi_1 e^{2V} \overline{\Phi}_1 + \alpha\Phi_2 e^{-2V}\overline{\Phi}_2 + 2CV \right]
+ \left\{ \frac{\Lambda}{4} \int d^2\theta \, (\Phi_1 \Phi_2) + {\rm h.c.} \right\}, \label{NA2}
\end{align}
where $\tilde{\Phi}\equiv \Phi+\Phi_1$ is the same as that in Eq.\,(\ref{GA3}). 
Note that this field redefinition is consistent with the U(1) symmetry
since $\Phi$ and $\Phi_1$ have the same U(1) charges. 

The variation with respect to $V$ yields a constraint equation,
\begin{align}
\tilde{\Phi} e^{2V} \overline{\tilde{\Phi}} - \Phi_1e^{2V} \overline{\Phi}_1 - \alpha\Phi_2 e^{-2V} \overline{\Phi}_2 + C = 0. \label{const}
\end{align}
Note that we need to fix the $U(1)^\C$ gauge redundancy, 
by which we can set one of the superfields as a constant. 
From the lowest component of this superfield equation in Wess-Zumino gauge, 
we obtain the following constraint on scalar fields:
\begin{align}
|\tilde{\phi}|^2 - |\phi_1|^2 - \alpha|\phi_2|^2 + C = 0.
\end{align}
For the consistency of this equation, 
at least one of the scalar field have to be nonzero.
When $\phi_1$ is nonzero, 
it is convenient to define the following gauge invariant superfields
\beq
X = \frac{\tilde{\Phi} \ }{\Phi_1}, \hs{10} Y = \Phi_1 \Phi_2. 
\eeq
Similarly, when $\tilde{\phi}$ or $\phi_1$ is nonzero, 
we can define gauge invariant superfields 
which are related to $(X, Y)$ by a field redefinition.  
For the moment, we assume that $\Phi_1$ is nonzero. 

Let us discuss solutions for Eq.\,(\ref{const}). 
The formal solutions of Eq.\,(\ref{const}) are given by
\begin{align}
e^{2V} 
\equiv G_{\pm} 
= \frac{1}{2|\Phi_1|^2} \frac{C \pm f \ }{1-|X|^2},
\label{defG}
\end{align}
where we have defined the function $f$ as
\beq
f = \sqrt{C^2 - 4 \alpha |Y|^2 (1-|X|^2)}. 
\eeq
Substituting into the K\"ahler potential, we obtain
\beq
K &=& \tilde{\Phi} e^{2V} \overline{\tilde{\Phi}} - \Phi_1 e^{2V} \overline{\Phi}_1 + \alpha\Phi_2 e^{-2V} \overline{\Phi}_2 + 2CV  \phantom{\bigg(} \notag \\
&=& \mp f + C \log \frac{C \pm f \ }{1-|X|^2} - C \log 2|\Phi_1|^2.
\label{Kahler}
\eeq
Note that the last term is unphysical since it can be eliminated by a K\"ahler transformation. 
Since the K\"ahler potential is written in terms of $|X|^2$ and $|Y|^2$, 
it has a U(1)$^2$ holomorphic isometry which remains after gauging one U(1) symmetry among the U(1)$^3$ symmetry of the ungauged action \eqref{component}.

The K\"ahler metric for $X$ and $Y$ is given by 
\beq
 K_{i \bar j} =
\ba{cc}
K_{X\overline{X}} & K_{X\overline{Y}} \\
K_{Y\overline{X}} & K_{Y\overline{Y}}
\ea
=
\pm
\frac{\alpha}{f}
\ba{cc}
H_{\pm} - |Y|^2 & -\overline{X}Y \\
-X \overline{Y} & 1-|X|^2
\ea, 
\eeq
where the function $H_{\pm}$ is given by
\beq
H_{\pm} = \frac{1}{4\alpha} \left( \frac{C \pm f \ }{1-|X|^2} \right)^2.
\eeq
To find out the condition for the positive definiteness of the K\"ahler metric, 
let us consider the following pair of linearly independent vectors 
\beq
\xi = \frac{\p}{\p Y}, \hs{10} 
\xi' = (1-|X|^2) \frac{\p}{\p X} + \overline{X} Y \frac{\p}{\p Y}. 
\eeq
Since they are mutually orthogonal, 
the K\"ahler metric is positive definite if both of the following norms of the vectors are positive:
\beq
 K_{i\bar j} \xi^i \bar \xi^{\bar j} = \pm \frac{\alpha}{f}(1-|X|^2), \hs{10}
 K_{i\bar j} \xi^{\prime i} \bar \xi^{\prime\bar j} = \frac{C \pm f}{2}.
\eeq
In addition, the solution for the auxiliary vector superfield Eq.\,\eqref{defG} also has to be positive, 
i.e. 
\beq
G_{\pm} = \pm \frac{\alpha f}{|\Phi_1|^2} \frac{K_{i\bar j} \xi^{\prime i} \bar \xi^{\prime\bar j}}{K_{i\bar j} \xi^i \bar \xi^j} > 0.
\eeq
These conditions are satisfied if and only if 
we choose $G_+$ and the following conditions are satisfied:
\beq
\alpha > 0, \hs{10} 
C > 0, \hs{10} 
|X|^2 < 1, \hs{10} 
|Y|^2 \leq \frac{C^2}{4\alpha} \frac{1}{1-|X|^2}.
\label{conditions}
\eeq
The kinetic coefficients are positive around the region satisfying the conditions. 
In the discussion above, we have assumed that $\Phi_1$ is nonzero. 
If $\Phi_1=0$, the condition \eqref{conditions} is not satisfied 
$\Phi_1 \approx 0$ corresponds to the region $X \approx \infty$. 
Therefore, $\Phi_1$ has to be nonzero in the physically consistent situation. 

Let us discuss a particular region satisfying the conditions realizing a stable system. 
We focus on the field region around $X \simeq Y \simeq 0$ 
($\tilde{\Phi}\simeq \Phi_2\simeq0$),  
where the K\"ahler potential~(\ref{Kahler}) is approximately given by
\begin{align}
K \simeq C |X|^2 + \frac{\alpha}{C} |Y|^2.
\end{align}
Thus, both of superfields are not ghost-like, and the instability is completely removed.
However, one needs also to check whether the resultant action is compatible with the conditions. 
In particular, the remaining superfields have a superpotential term, 
which gives rise to a scalar potential. 
In the following, we show two concrete regions satisfying the conditions 
and also consistent with the vacuum determined by the scalar potential.

Let us make a comment on the geometry of our model.
The resultant target space $M$ is a 
certain fiber bundle over a hyperbolic space,
\begin{align}
 M \simeq D \ltimes {SU(1,1) \over U(1)}
\end{align}
where $F \ltimes B$ denotes a fiber bundle over a base $B$ with 
a fiber $F$.
The base is parameterized by $X$,
while the fiber $D$ is a disk parameterized by $Y$ having 
a range determined by $X$ 
in Eq.~(\ref{conditions}).

\subsubsection{The second model}\label{models}

It is worth noting that the case with $\alpha<0,\ C<0$ 
has the same structure as the case with $\alpha>0,\ C>0$
if we exchange $\tilde \Phi \leftrightarrow \Phi_1$ and 
flip the sign of the K\"ahler potential $K \to - K$. 
Therefore, for $\alpha<0,\ C<0$, the K\"ahler metric is positive definite, 
i.e. the model is ghost-free. 
Since we have flipped the sign of the K\"ahler potential, 
the corresponding Lagrangian has to have the negative kinetic term for $\Phi$ 
\begin{align}
{\cal L}''=\int d^4 \theta \left[ - \Phi e^{2V} \overline{\Phi} + \frac{\tilde{\alpha}}{\Lambda^2} \Big( D^2(\Phi e^{2V}) \Big) e^{-2V} \Big( \overline{D}^2 ( \overline{\Phi} e^{2V}) \Big) + 2\tilde{C}V \right], \label{NNA}
\end{align}
where $\tilde{\alpha}=-\alpha>0$ and $\tilde{C}=-C>0$. 
This Lagrangian is similar to that in Eq.~(\ref{NA}), 
but the signs of the first terms are opposite to each other. 
This is not difficult to understand by the following reason: 
From the procedure in Sec.\,\ref{rev}, 
we find that $\tilde{\Phi}$ and $\Phi_1$ always have the opposite sign. 
The one with a negative sign is regarded as the ghost mode. 
In the case with (\ref{NNA}), we can identify $\tilde{\Phi}$ as the ghost, 
and remove it by our mechanism. 
The effective K\"ahler metric around 
$\tilde{X} \equiv \Phi_1/\tilde{\Phi} \simeq 0$, $\tilde{Y} \equiv \tilde{\Phi} \Phi_2 \simeq 0$ is given by
\begin{align}
K \simeq \tilde{C} |\tilde{X}|^2 + \frac{\tilde{\alpha}}{\tilde{C}} |\tilde{Y}|^2.
\end{align} 
This system clearly has no ghost modes, as is the case with $\alpha>0,\ C>0$ discussed above. 
Although it seems that there is no difference between these cases, 
the system has a completely different vacuum structure 
because of the difference of the superpotentials, 
as will be discussed in the next section. 

\section{Behaviour of dynamical F-term superfield and SUSY breaking/preserving vacua}\label{DF-term}
In this section, we consider the structure of the vacuum in our model. 

\subsection{The first model: SUSY breaking vacuum}
In the previous section, we have seen that there is no ghost mode 
if the conditions \eqref{conditions} are satisfied. 
In this case, the superpotential is linear in the chiral superfield $Y= \Phi_1 \Phi_2$
\begin{align}
W= \frac{\Lambda}{4} Y .
\end{align}
Therefore, the supersymmetry is spontaneously broken due to the nonzero values of 
the F-terms of $Y$ and the F-term scalar potential
\begin{align}
F^{Y} & = -\frac{\Lambda}{4}, \hs{10} 
V=\frac{C\Lambda^2}{16\alpha}.\label{V1}
\end{align}

It is worth noting that $\Phi_2$ plays a role of a SUSY breaking field as the Polonyi model. This means that the ``dynamical F-term" superfield $\Phi_2$ breaks SUSY spontaneously, and the order of SUSY breaking is determined by the FI parameter $C$ and the cut-off $\Lambda$. 

Here,
let us make a comment on SUSY breaking in different supermultiplets 
with higher derivative term.
It is known that a complex linear superfield with its higher-derivative term is dual to a chiral and nilpotent-chiral superfield~\cite{Farakos:2013zsa,Farakos:2014iwa,Farakos:2015vba} (see also Refs.~\cite{Kuzenko:2011ti,Kuzenko:2015uca}). The nilpotent-chiral superfield also spontaneously breaks SUSY, as $\Phi_2$ in our case. The relation between the higher-derivative extension of complex linear and chiral superfield discussed here is quite interesting. 
To the best of our knowledge, our model gives the first example of 
such SUSY breaking by a higher derivative term in chiral multiplets.
We will investigate extensions of this SUSY breaking mechanism elsewhere.

\subsection{The second model: SUSY preserving vacuum}
On the other hand, in the model~(\ref{NNA}), 
the structure of the superpotential is essentially different. 
In terms of $\tilde{X} \equiv \Phi_1/\tilde{\Phi}$ and $\tilde{Y} \equiv \tilde{\Phi} \Phi_2$
the superpotential is given by
\begin{align}
W=\frac{\Lambda}{4} \tilde{X} \tilde{Y}.
\end{align}
The F-term of $\tilde{X}$ and $\tilde{Y}$ are 
\begin{align}
\overline{F}^{\tilde{X}}&=-\frac{\Lambda}{4\tilde{C}} \tilde{Y},\\
\overline{F}^{\tilde{Y}}&=-\frac{\Lambda \tilde{C}}{4 \tilde \alpha} \tilde{X}.
\end{align}
Since the scalar potential becomes
\begin{align}
 V \sim \frac{\Lambda^2}{16 \tilde{\alpha}} \left( \tilde{C} |\tilde{X}|^2 + \frac{\tilde{\alpha}}{\tilde{C}} |\tilde{Y}|^2 \right),
\end{align}
the vacuum $\tilde{X}=\tilde{Y}=0$ ($\Phi_1=\Phi_2=0$) is stable and 
the F-terms do not have vacuum expectation values. 
Therefore, with the Lagrangian~(\ref{NNA}), SUSY is preserved at the vacuum.

It is quite interesting that we can realize both SUSY preserving or breaking vacuum from the almost the same systems (\ref{NA}) and (\ref{NNA}). In the case~(\ref{NA}), the higher-derivative term induces the ghost, and after removing it, we obtain the SUSY breaking vacuum with a cosmological constant, as shown in Eq.~(\ref{V1}). On the other hand, when we start with a ghost-like superfield with its higher-derivative term~(\ref{NNA}), we finally obtain the model with a SUSY preserving vacuum. We will investigate this feature in more detail elsewhere.

Note that SUSY breaking in our model is different from that in SUSY ghost condensation~\cite{Khoury:2010gb}, in which the violation of time translation invariance occurs. In our model, SUSY is broken in a Lorentz invariant manner.

\section{Summary and discussion}\label{concl}
In this paper, we have proposed a new mechanism to construct ghost-free higher-derivative models within global SUSY. The important notion is a non-dynamical gauge superfield, which ``eats" the ghost mode in the system. 
We have illustrated our mechanism with an example shown in Sec.~\ref{rev}, which has a ghost superfield. As shown in Sec.~\ref{exo}, the ghost mode can be removed thanks to the non-dynamical gauge superfield. It has been shown that, independently of the sign of a kinetic term of $\Phi$, the higher-derivative system has one normal and one ghost mode, and we can remove a ghost superfield in the both cases. Interestingly enough, however, the resultant systems after removing a ghost are completely different from each other as discussed in Sec.~\ref{DF-term}. In particular, the vacuum structures are different: One gives a SUSY breaking vacuum, and the other gives a SUSY preserving vacuum. 
The former is the first example of SUSY breaking induced by 
a higher derivative term in chiral superfields.
One of the most interesting features is 
that 
because of the higher derivative term including 
space-time derivative on 
the F-term in the original chiral superfield $\Phi$, 
the F-term becomes dynamical 
and resides in the lowest component of 
the chiral superfield $\Phi_2$. 

The remaining question is the physical meaning of 
the a propagating F-term field $F$. 
As mentioned, the F-term is now in the lowest component of 
the chiral superfield $\Phi_2$,
and so 
the structures of the SUSY multiplets 
are completely different from the 
original multiplets in the absence of 
the higher derivative term $(\alpha=0)$. 
One of consequences of the propagating F-term 
is, as shown in this paper, the existence of 
the U(1) conserved charge associated with the phase of $F$.
It should be important to study more consequences 
for instance 
the structure of SUSY algebra and so on.
The physical meaning of the SUSY breaking vacuum 
is unclear.

We have illustrated our mechanism of eliminating a ghost 
and introducing a dynamical auxiliary field in the simplest example. 
One of straightforward extensions is 
multiple chiral superfields. 
When the superderivative $D^2$ acts on $n$ chiral superfields, 
there will be at most $n$ ghost fields. 
Therefore, we need at least $U(1)^n$ gauging. 
With this regards, a higher derivative 
${\mathbb C}P^1$ model in Ref.~\cite{Bergshoeff:1984wb} 
contains two chiral superfields and only one U(1) gauge field. 
Consequently only one of two ghosts would be removed 
but the rest would remain, and so the theory is pathological.
Another possible extension is a non-Abelian extension.
For instance, if the original Lagrangian contains 
an $N$ by $N$ matrix chiral superfield, 
a ``non-Abelian'' (matrix) ghost superfield appears.
This could be removed by $U(N)$ gauging.
It is desired to construct a general framework by classifying 
how many ghosts non-gauged theories have, 
and which gauging (U(1), $U(N)$ or other gauge groups) 
can remove those ghosts. 
In other words, 
it would be very important to construct 
``generalized Nambu-Goldstone theorem'' including 
ghosts and ``generalized Higgs mechanism.''

In the language of the K\"ahler geometry,
eliminating vector superfield is known as the K\"ahler quotient.
Usually this has been studied very well for positive norm metrics.
Generalizing the K\"ahler quotient to include negative norm metrics  
 should be a key point to understand the whole theory geometrically.

Our vector superfield is a non-dynamical and auxiliary field 
behaving as a Lagrange multiplier.
If we add a kinetic term of the vector superfield, 
a gauge field absorbing a ghost will have a tachyonic mass, 
so still having the instability 
(our case can be understood as sending away such the tachyonic 
mass to infinity). 
The auxiliary field formulation of nonlinear sigma models 
in lower (1+1 or 2+1) dimensions 
often results in a kinetic term of the gauge field 
by the quantum effect, 
as can be explicitly shown in the large-$N$ limit. 
If it was the same for our model in lower dimensions, 
there would be the quantum mechanically induced 
instability which is absent at the tree level.
It is very interesting to study whether the instability exists or not 
in the quantization of dimensionally reduced model 
(or even the 3+1 dimensional theory as a cut-off theory).

In the formulation of (supersymmetric) ${\mathbb C}P^n$ model in terms of 
an auxiliary U(1) gauge field (vector superfield),
a vortex (flux tube) carrying the U(1) gauge magnetic field is 
nothing but a  ${\mathbb C}P^n$ sigma model lump (instanton). 
Whether our model admits such a lump and its stability (if it exists) 
are an interesting question.

As a non-dynamical gauge superfield and a propagating F-term regard, 
it is worth mentioning the similarity between our mechanism and a compensator in conformal SUGRA. In conformal SUGRA, we usually introduce a ghost-like superfield called a compensator~\cite{Kaku:1978ea,Kugo:1982cu}.  The compensator is removed by the conformal gauge degrees of freedom, and finally the system does not have any ghost-like mode. The gauge fields of conformal symmetries are non-propagating as the gauge superfield in our mechanism. From this viewpoint, the ghost mode in our model is similar to the compensator, and the non-dynamical gauge superfield to the conformal gauge fields. In addition, the propagating auxiliary fields have been discussed in the context of the higher-curvature SUGRA model~\cite{Cecotti:1987sa,Farakos:2013cqa,Ketov:2013sfa,Ferrara:2013wka,Ketov:2013dfa}, where auxiliary fields in the gravity multiplet obtain the kinetic term due to higher-derivative terms of the gravity multiplet. Therefore, the presence of dynamical auxiliary fields may not be problematic.

Finally, the coupling of our model to SUGRA 
should be interesting for applications to cosmology 
such as inflationary models.

\section*{Acknowledgement}
M. N. would like to thank Shin Sasaki for bringing his attention  
to SUSY higher-derivative models.
Y.Y. would like to thank Shuntaro Aoki and Tomomi Ihara for useful discussion on issues of SUSY higher-derivatives and ghosts.
This work is supported by the Ministry of Education,
Culture, Sports, Science (MEXT)-Supported Program for the Strategic
Research Foundation at Private Universities ``Topological Science''
(Grant No.~S1511006). 
The work of M.~N.~is also
supported in part by the Japan Society for the Promotion of Science
(JSPS) Grant-in-Aid for Scientific Research (KAKENHI Grant
No.~25400268) and by 
a Grant-in-Aid for
Scientific Research on Innovative Areas ``Topological Materials
Science'' (KAKENHI Grant No.~15H05855) and ``Nuclear Matter in Neutron
Stars Investigated by Experiments and Astronomical Observations''
(KAKENHI Grant No.~15H00841) from the MEXT of Japan.


\begin{thebibliography}{99}
\bibitem{Woodard:2006nt} 
  R.~P.~Woodard,
  ``Avoiding dark energy with 1/r modifications of gravity,''
  Lect.\ Notes Phys.\  {\bf 720}, 403 (2007)
  [astro-ph/0601672].
\bibitem{Horndeski:1974wa} 
  G.~W.~Horndeski,
  ``Second-order scalar-tensor field equations in a four-dimensional space,''
  Int.\ J.\ Theor.\ Phys.\  {\bf 10}, 363 (1974).
\bibitem{Kobayashi:2011nu} 
  T.~Kobayashi, M.~Yamaguchi and J.~Yokoyama,
  ``Generalized G-inflation: Inflation with the most general second-order field equations,''
  Prog.\ Theor.\ Phys.\  {\bf 126}, 511 (2011)
   [arXiv:1105.5723 [hep-th]].
\bibitem{Charmousis:2011bf} 
  C.~Charmousis, E.~J.~Copeland, A.~Padilla and P.~M.~Saffin,
  ``General second order scalar-tensor theory, self tuning, and the Fab Four,''
  Phys.\ Rev.\ Lett.\  {\bf 108}, 051101 (2012)
  [arXiv:1106.2000 [hep-th]].

\bibitem{Bagger:1996wp} 
  J.~Bagger and A.~Galperin,
  ``A New Goldstone multiplet for partially broken supersymmetry,''
  Phys.\ Rev.\ D {\bf 55}, 1091 (1997)
    [hep-th/9608177].
\bibitem{Bagger:1997pi} 
  J.~Bagger and A.~Galperin,
  ``The Tensor Goldstone multiplet for partially broken supersymmetry,''
  Phys.\ Lett.\ B {\bf 412}, 296 (1997)
   [hep-th/9707061].
\bibitem{Rocek:1997hi} 
  M.~Rocek and A.~A.~Tseytlin,
  ``Partial breaking of global D = 4 supersymmetry, constrained superfields, and three-brane actions,''
  Phys.\ Rev.\ D {\bf 59}, 106001 (1999)
   [hep-th/9811232].
  \bibitem{Kuzenko:2002vk} 
  S.~M.~Kuzenko and S.~A.~McCarthy,
  ``Nonlinear selfduality and supergravity,''
  JHEP {\bf 0302}, 038 (2003)
  [hep-th/0212039].
  \bibitem{Kuzenko:2005wh} 
  S.~M.~Kuzenko and S.~A.~McCarthy,
  ``On the component structure of N=1 supersymmetric nonlinear electrodynamics,''
  JHEP {\bf 0505}, 012 (2005)
  [hep-th/0501172].
  \bibitem{Farakos:2012je} 
  F.~Farakos, C.~Germani, A.~Kehagias and E.~N.~Saridakis,
  ``A New Class of Four-Dimensional N=1 Supergravity with Non-minimal Derivative Couplings,''
  JHEP {\bf 1205}, 050 (2012)
    [arXiv:1202.3780 [hep-th]].
  \bibitem{Koehn:2012ar} 
  M.~Koehn, J.~L.~Lehners and B.~A.~Ovrut,
  ``Higher-Derivative Chiral Superfield Actions Coupled to N=1 Supergravity,''
  Phys.\ Rev.\ D {\bf 86}, 085019 (2012)
  [arXiv:1207.3798 [hep-th]].
  \bibitem{Abe:2015nxa} 
  H.~Abe, Y.~Sakamura and Y.~Yamada,
  ``Matter coupled Dirac-Born-Infeld action in four-dimensional N=1 conformal supergravity,''
  Phys.\ Rev.\ D {\bf 92}, no. 2, 025017 (2015)
    [arXiv:1504.01221 [hep-th]].
  \bibitem{Aoki:2016cnw} 
  S.~Aoki and Y.~Yamada,
  ``DBI action of real linear superfield in 4D ${\cal N}=1$ conformal supergravity,''
  JHEP06(2016)168
   [arXiv:1603.06770 [hep-th]].



\bibitem{Khoury:2010gb} 
  J.~Khoury, J.~L.~Lehners and B.~Ovrut,
  ``Supersymmetric P(X,$\phi$) and the Ghost Condensate,''
  Phys.\ Rev.\ D {\bf 83}, 125031 (2011)
   [arXiv:1012.3748 [hep-th]].
\bibitem{Farakos:2013zya} 
  F.~Farakos, C.~Germani and A.~Kehagias,
  ``On ghost-free supersymmetric galileons,''
  JHEP {\bf 1311}, 045 (2013)
    [arXiv:1306.2961 [hep-th]].
  \bibitem{Sasaki:2012ka} 
  S.~Sasaki, M.~Yamaguchi and D.~Yokoyama,
  ``Supersymmetric DBI inflation,''
  Phys.\ Lett.\ B {\bf 718}, 1 (2012)
    [arXiv:1205.1353 [hep-th]].
  \bibitem{Koehn:2012np} 
  M.~Koehn, J.~L.~Lehners and B.~A.~Ovrut,
  ``DBI Inflation in N=1 Supergravity,''
  Phys.\ Rev.\ D {\bf 86}, 123510 (2012)
    [arXiv:1208.0752 [hep-th]].
  \bibitem{Gwyn:2014wna} 
  R.~Gwyn and J.~L.~Lehners,
  ``Non-Canonical Inflation in Supergravity,''
  JHEP {\bf 1405}, 050 (2014)
    [arXiv:1402.5120 [hep-th]].
  \bibitem{Aoki:2014pna} 
  S.~Aoki and Y.~Yamada,
  ``Inflation in supergravity without Kahler potential,''
  Phys.\ Rev.\ D {\bf 90}, no. 12, 127701 (2014)
  [arXiv:1409.4183 [hep-th]].
  \bibitem{Aoki:2015eba} 
  S.~Aoki and Y.~Yamada,
  ``Impacts of supersymmetric higher derivative terms on inflation models in supergravity,''
  JCAP {\bf 1507}, no. 07, 020 (2015)
    [arXiv:1504.07023 [hep-th]].
  \bibitem{Abe:2015fha} 
  H.~Abe, Y.~Sakamura and Y.~Yamada,
  ``Massive vector multiplet inflation with Dirac-Born-Infeld type action,''
  Phys.\ Rev.\ D {\bf 91}, no. 12, 125042 (2015)
   [arXiv:1505.02235 [hep-th]].
  
\bibitem{Adam:2011hj} 
  C.~Adam, J.~M.~Queiruga, J.~Sanchez-Guillen and A.~Wereszczynski,
  ``N=1 supersymmetric extension of the baby Skyrme model,''
  Phys.\ Rev.\ D {\bf 84}, 025008 (2011)
   [arXiv:1105.1168 [hep-th]].

\bibitem{Adam:2013awa} 
  C.~Adam, J.~M.~Queiruga, J.~Sanchez-Guillen and A.~Wereszczynski,
  ``Extended Supersymmetry and BPS solutions in baby Skyrme models,''
  JHEP {\bf 1305}, 108 (2013)
   [arXiv:1304.0774 [hep-th]].

\bibitem{Nitta:2014pwa} 
  M.~Nitta and S.~Sasaki,
  ``BPS States in Supersymmetric Chiral Models with Higher Derivative Terms,''
  Phys.\ Rev.\ D {\bf 90}, no. 10, 105001 (2014)
  [arXiv:1406.7647 [hep-th]].
 
\bibitem{Bolognesi:2014ova} 
  S.~Bolognesi and W.~Zakrzewski,
  ``Baby Skyrme Model, Near-BPS Approximations and Supersymmetric Extensions,''
  Phys.\ Rev.\ D {\bf 91}, no. 4, 045034 (2015)
  [arXiv:1407.3140 [hep-th]].

\bibitem{Queiruga:2016jqu} 
  J.~M.~Queiruga,
  ``Baby Skyrme model and fermionic zero modes,''
  arXiv:1606.02869 [hep-th].

\bibitem{Nitta:2015uba} 
  M.~Nitta and S.~Sasaki,
  ``Classifying BPS States in Supersymmetric Gauge Theories Coupled to Higher Derivative Chiral Models,''
  Phys.\ Rev.\ D {\bf 91}, 125025 (2015)
   [arXiv:1504.08123 [hep-th]].
 

\bibitem{Buchbinder:1994iw} 
  I.~L.~Buchbinder, S.~Kuzenko and Z.~Yarevskaya,
  ``Supersymmetric effective potential: Superfield approach,''
  Nucl.\ Phys.\ B {\bf 411}, 665 (1994);

  I.~L.~Buchbinder and S.~M.~Kuzenko,
  ``Ideas and methods of supersymmetry and supergravity: Or a walk through superspace,''
  Bristol, UK: IOP (1998) 656 p; 
  
  S.~M.~Kuzenko and S.~J.~Tyler,
  ``The one-loop effective potential of the Wess-Zumino model revisited,''
  arXiv:1407.5270 [hep-th].
\bibitem{Banin:2006db} 
  A.~T.~Banin, I.~L.~Buchbinder and N.~G.~Pletnev,
  ``On quantum properties of the four-dimensional generic chiral superfield model,''
  Phys.\ Rev.\ D {\bf 74}, 045010 (2006)
  [hep-th/0606242].

\bibitem{Nemeschansky:1984cd} 
  D.~Nemeschansky and R.~Rohm,
  ``Anomaly Constraints On Supersymmetric Effective Lagrangians,''
  Nucl.\ Phys.\ B {\bf 249}, 157 (1985).
  
\bibitem{Gates:1995fx} 
  S.~J.~Gates, Jr.,
  ``Why auxiliary fields matter: The Strange case of the 4-D, N=1 supersymmetric QCD effective action,''
  Phys.\ Lett.\ B {\bf 365}, 132 (1996)
  [hep-th/9508153], 
  ``Why auxiliary fields matter: The strange case of the 4-D, N=1 supersymmetric QCD effective action. 2.,''
  Nucl.\ Phys.\ B {\bf 485}, 145 (1997)
  [hep-th/9606109].
  
  
\bibitem{Gates:2000rp} 
  S.~J.~Gates, Jr., M.~T.~Grisaru, M.~E.~Knutt and S.~Penati,
  ``The Superspace WZNW action for 4-D, N=1 supersymmetric QCD,''
  Phys.\ Lett.\ B {\bf 503}, 349 (2001)
  [hep-ph/0012301];
  S.~J.~Gates, Jr., M.~T.~Grisaru, M.~E.~Knutt, S.~Penati and H.~Suzuki,
  ``Supersymmetric gauge anomaly with general homotopic paths,''
  Nucl.\ Phys.\ B {\bf 596}, 315 (2001)
  [hep-th/0009192];
  
   S.~J.~Gates, Jr., M.~T.~Grisaru and S.~Penati,
  ``Holomorphy, minimal homotopy and the 4-D, N=1 supersymmetric Bardeen-Gross-Jackiw anomaly,''
  Phys.\ Lett.\ B {\bf 481}, 397 (2000)
  [hep-th/0002045].
 
\bibitem{Nitta:2001rh} 
  M.~Nitta,
  ``A Note on supersymmetric WZW term in four dimensions,''
  Mod.\ Phys.\ Lett.\ A {\bf 15}, 2327 (2000)
    [hep-th/0101166].

\bibitem{Nitta:2014fca} 
  M.~Nitta and S.~Sasaki,
  ``Higher Derivative Corrections to Manifestly Supersymmetric Nonlinear Realizations,''
  Phys.\ Rev.\ D {\bf 90}, no. 10, 105002 (2014)
  [arXiv:1408.4210 [hep-th]].
 
\bibitem{Eto:2012qda} 
  M.~Eto, T.~Fujimori, M.~Nitta, K.~Ohashi and N.~Sakai,
  ``Higher Derivative Corrections to Non-Abelian Vortex Effective Theory,''
  Prog.\ Theor.\ Phys.\  {\bf 128}, 67 (2012)
    [arXiv:1204.0773 [hep-th]].


\bibitem{Adam:2011gc} 
  C.~Adam, J.~M.~Queiruga, J.~Sanchez-Guillen and A.~Wereszczynski,
  ``Supersymmetric K field theories and defect structures,''
  Phys.\ Rev.\ D {\bf 84}, 065032 (2011)
  [arXiv:1107.4370 [hep-th]].
 
\bibitem{Adam:2012md} 
  C.~Adam, J.~M.~Queiruga, J.~Sanchez-Guillen and A.~Wereszczynski,
  ``BPS bounds in supersymmetric extensions of K field theories,''
  Phys.\ Rev.\ D {\bf 86}, 105009 (2012)
  [arXiv:1209.6060 [hep-th]].
 
\bibitem{Freyhult:2003zb} 
  L.~Freyhult,
  ``The Supersymmetric extension of the Faddeev model,''
  Nucl.\ Phys.\ B {\bf 681}, 65 (2004)
    [hep-th/0310261].

\bibitem{Bergshoeff:1984wb} 
  E.~A.~Bergshoeff, R.~I.~Nepomechie and H.~J.~Schnitzer,
  ``Supersymmetric Skyrmions in Four-dimensions,''
  Nucl.\ Phys.\ B {\bf 249}, 93 (1985).
\bibitem{Gudnason:2015ryh} 
  S.~B.~Gudnason, M.~Nitta and S.~Sasaki,
  ``A supersymmetric Skyrme model,''
  JHEP {\bf 1602}, 074 (2016)
   [arXiv:1512.07557 [hep-th]].
 
\bibitem{Gomes:2009ev} 
  M.~Gomes, J.~R.~Nascimento, A.~Y.~Petrov and A.~J.~da Silva,
  ``On the effective potential in higher-derivative superfield theories,''
  Phys.\ Lett.\ B {\bf 682}, 229 (2009)
  [arXiv:0908.0900 [hep-th]].
  
\bibitem{Gama:2011ws} 
  F.~S.~Gama, M.~Gomes, J.~R.~Nascimento, A.~Y.~Petrov and A.~J.~da Silva,
  ``On the higher-derivative supersymmetric gauge theory,''
  Phys.\ Rev.\ D {\bf 84}, 045001 (2011)
  [arXiv:1101.0724 [hep-th]].
  
\bibitem{Gama:2014fca} 
  F.~S.~Gama, M.~Gomes, J.~R.~Nascimento, A.~Y.~Petrov and A.~J.~da Silva,
  ``On the one-loop effective potential in the higher-derivative four-dimensional chiral superfield theory with a nonconventional kinetic term,''
  Phys.\ Lett.\ B {\bf 733}, 247 (2014)
  [arXiv:1401.5414 [hep-th]].
\bibitem{Kimura:2016irk} 
  T.~Kimura, A.~Mazumdar, T.~Noumi and M.~Yamaguchi,
  ``Nonlocal N=1 Supersymmetry,''
  arXiv:1608.01652 [hep-th].  
\bibitem{Antoniadis:2007xc} 
  I.~Antoniadis, E.~Dudas and D.~M.~Ghilencea,
  ``Supersymmetric Models with Higher Dimensional Operators,''
  JHEP {\bf 0803}, 045 (2008)
    [arXiv:0708.0383 [hep-th]].
\bibitem{Dudas:2015vka} 
  E.~Dudas and D.~M.~Ghilencea,
  ``Effective operators in SUSY, superfield constraints and searches for a UV completion,''
  JHEP {\bf 1506}, 124 (2015)
  [arXiv:1503.08319 [hep-th]].
 
 
 
  
  \bibitem{Cecotti:1987sa} 
  S.~Cecotti,
  ``Higher Derivative Supergravity Is Equivalent To Standard Supergravity Coupled To Matter. 1.,''
  Phys.\ Lett.\ B {\bf 190}, 86 (1987).
\bibitem{Farakos:2013cqa} 
  F.~Farakos, A.~Kehagias and A.~Riotto,
  ``On the Starobinsky Model of Inflation from Supergravity,''
  Nucl.\ Phys.\ B {\bf 876}, 187 (2013)
  [arXiv:1307.1137 [hep-th]].
\bibitem{Ketov:2013sfa} 
  S.~V.~Ketov,
  ``On the supersymmetrization of inflation in f(R) gravity,''
  PTEP {\bf 2013}, 123B04 (2013)
   [arXiv:1309.0293 [hep-th]].
\bibitem{Ferrara:2013wka} 
  S.~Ferrara, R.~Kallosh and A.~Van Proeyen,
  ``On the Supersymmetric Completion of $R+R^2$ Gravity and Cosmology,''
  JHEP {\bf 1311}, 134 (2013)
   [arXiv:1309.4052 [hep-th]].
\bibitem{Ketov:2013dfa} 
  S.~V.~Ketov and T.~Terada,
  ``Old-minimal supergravity models of inflation,''
  JHEP {\bf 1312}, 040 (2013)
   [arXiv:1309.7494 [hep-th]].
  
\bibitem{D'Adda:1978uc} 
  A.~D'Adda, M.~Luscher and P.~Di Vecchia,
  ``A 1/n Expandable Series of Nonlinear Sigma Models with Instantons,''
  Nucl.\ Phys.\ B {\bf 146}, 63 (1978).
   
\bibitem{Higashijima:1999ki} 
  K.~Higashijima and M.~Nitta,
  ``Supersymmetric nonlinear sigma models as gauge theories,''
  Prog.\ Theor.\ Phys.\  {\bf 103}, 635 (2000)
  [hep-th/9911139]; 
  M.~Nitta,
  ``Auxiliary field methods in supersymmetric nonlinear sigma models,''
  Nucl.\ Phys.\ B {\bf 711}, 133 (2005)
    [hep-th/0312025].
  


 


 
 
  
\bibitem{Farakos:2013zsa} 
  F.~Farakos, S.~Ferrara, A.~Kehagias and M.~Porrati,
  ``Supersymmetry Breaking by Higher Dimension Operators,''
  Nucl.\ Phys.\ B {\bf 879}, 348 (2014)
  [arXiv:1309.1476 [hep-th]].
\bibitem{Farakos:2014iwa} 
  F.~Farakos and R.~von Unge,
  ``Complex Linear Effective Theory and Supersymmetry Breaking Vacua,''
  Phys.\ Rev.\ D {\bf 91}, no. 4, 045024 (2015)
  [arXiv:1403.0935 [hep-th]].
\bibitem{Farakos:2015vba} 
  F.~Farakos, O.~Hul\'ik, P.~Ko\v{c}\'i and R.~von Unge,
  ``Non-minimal scalar multiplets, supersymmetry breaking and dualities,''
  JHEP {\bf 1509}, 177 (2015)
   [arXiv:1507.01885 [hep-th]].
\bibitem{Kuzenko:2011ti} 
  S.~M.~Kuzenko and S.~J.~Tyler,
  ``Complex linear superfield as a model for Goldstino,''
  JHEP {\bf 1104}, 057 (2011)
    [arXiv:1102.3042 [hep-th]].
\bibitem{Kuzenko:2015uca} 
  S.~M.~Kuzenko and S.~J.~Tyler,
  ``Comments on the complex linear Goldstino superfield,''
  arXiv:1507.04593 [hep-th].
\bibitem{Kaku:1978ea} 
  M.~Kaku and P.~K.~Townsend,
  ``Poincare Supergravity As Broken Superconformal Gravity,''
  Phys.\ Lett.\ B {\bf 76}, 54 (1978).
\bibitem{Kugo:1982cu} 
  T.~Kugo and S.~Uehara,
  ``Conformal and Poincare Tensor Calculi in $N=1$ Supergravity,''
  Nucl.\ Phys.\ B {\bf 226}, 49 (1983).

\end{thebibliography}
\end{document}